\documentclass[a4paper,11pt]{article}
\pdfoutput=1

\usepackage{jcappub}
\usepackage[T1]{fontenc}
\usepackage{sidecap}
\usepackage{subcaption}
\usepackage{booktabs}
\usepackage{array}
\usepackage{cleveref}
\usepackage{xcolor}
\usepackage{soul}
\usepackage{hyperref}
\hypersetup{colorlinks   = true,
            urlcolor     = blue,
            citecolor    = blue,
            linkcolor    = blue,
            menucolor    = blue,
            anchorcolor  = blue,
            filecolor    = blue}

\enlargethispage{\baselineskip}
\DeclareCaptionJustification{justified}{\justifying}
\usepackage[normalem]{ulem}

\usepackage{tikz,xcolor,hyperref}

\definecolor{lime}{HTML}{A6CE39}
\DeclareRobustCommand{\orcidicon}{
	\begin{tikzpicture}
	\draw[lime, fill=lime] (0,0) 
	circle [radius=0.16] 
	node[white] {{\fontfamily{qag}\selectfont \tiny ID}};
	\draw[white, fill=white] (-0.0625,0.095) 
	circle [radius=0.007];
	\end{tikzpicture}
	\hspace{-2mm}
}
\foreach \x in {A, ..., Z}{\expandafter\xdef\csname orcid\x\endcsname{\noexpand\href{https://orcid.org/\csname orcidauthor\x\endcsname}
			{\noexpand\orcidicon}}
}

\title{Novel constraints on fifth forces and ultralight dark sector with asteroidal data}

\author[a,b,c]{Yu-Dai Tsai\hspace{-3mm}\orcidA{}\hspace{-3mm},}
\author[d]{Youjia Wu\hspace{-3mm}\orcidB{}\hspace{-3mm},}
\author[e,f,g]{Sunny Vagnozzi\hspace{-3mm}\orcidC{}\hspace{-3mm},}
\author[h,i]{\newline and Luca Visinelli\hspace{-3mm}\orcidD{}\hspace{-3mm}}
\affiliation[a]{Department of Physics and Astronomy, University of California, Irvine, CA 92697-4575, USA}
\affiliation[b]{Fermi National Accelerator Laboratory (Fermilab), Batavia, IL 60510, USA}
\affiliation[c]{Kavli Institute for Cosmological Physics (KICP), University of Chicago, Chicago, IL 60637, USA}
\affiliation[d]{Leinweber Center for Theoretical Physics, Department of Physics, University of Michigan, Ann Arbor, MI 48109, USA}
\affiliation[e]{Department of Physics, University of Trento, Via Sommarive 14, 38123 Povo (TN), Italy}
\affiliation[f]{Trento Institute for Fundamental Physics and Applications (TIFPA)-INFN, Via Sommarive 14, 38123 Povo (TN), Italy}
\affiliation[g]{Kavli Institute for Cosmology (KICC), University of Cambridge, Madingley Road, Cambridge CB3 0HA, United Kingdom}
\affiliation[g]{Tsung-Dao Lee Institute (TDLI), 520 Shengrong Road, 201210 Shanghai, P.\ R.\ China}
\affiliation[h]{School of Physics and Astronomy, Shanghai Jiao Tong University, 800 Dongchuan Road, 200240 Shanghai, P.\ R.\ China}

\emailAdd{yt444@cornell.edu}
\emailAdd{youjiawu@umich.edu}
\emailAdd{sunny.vagnozzi@unitn.it}
\emailAdd{luca.visinelli@sjtu.edu.cn}

\abstract{We study for the first time the possibility of probing long-range fifth forces utilizing asteroid astrometric data, via the fifth force-induced orbital precession. We examine nine Near-Earth Object (NEO) asteroids whose orbital trajectories are accurately determined via optical and radar astrometry. Focusing on a Yukawa-type potential mediated by a new gauge field (dark photon) or a baryon-coupled scalar, we estimate the sensitivity reach for the fifth force coupling strength and mediator mass in the mass range $m \simeq (10^{-21}-10^{-15})\,{\rm eV}$, near the ``fuzzy'' dark matter region. Our estimated sensitivity is comparable to leading limits from equivalence principle tests, potentially exceeding these in a specific mass range. Furthermore, we set new constraints on Yukawa gravity.
The fifth force-induced precession increases with the orbital semi-major axis in the limit of a small mass $m$, motivating the study of objects further away from the Sun. We discuss future prospects for extending our study to more than a million asteroids, including NEOs, main-belt asteroids, Hildas, and Jupiter Trojans, as well as trans-Neptunian objects and exoplanets.}

\begin{document}
\maketitle
\flushbottom

\section{Introduction}

The study of precessions has revealed some of the deepest secrets of Nature. Most notably, the correct prediction for Mercury's precession rate from General Relativity (GR) is one of the theory's major successes~\cite{1859AnPar...5....1L, Einstein:1916vd,Corda:2021qeg}. The findings of the \textit{Muon} $g-2$ experiment measuring the muon anomalous precession frequency might hint at the existence of physics beyond the Standard Model (SM)~\cite{Muong-2:2006rrc,Davier:2019can,Abi:2021gix}.
New connections between microscopic physics and macroscopic planetary science can be established by studying the precessions of celestial objects, due to long-range forces mediated by (new) ultralight particles.

There are strong motivations to investigate the existence of new light, weakly-coupled degrees of freedom beyond the SM (BSM), which are generic features of string theory~\cite{Svrcek:2006yi, Arvanitaki:2009fg, Cicoli:2012sz, Visinelli:2018utg}, and are candidates for the dark matter (DM) and dark energy (DE)~\cite{Peccei:1987mm,Wetterich:1987fm,Ratra:1987rm,Wetterich:2002ic,Khoury:2003rn}. For example, ultralight (fuzzy) DM may play a significant role in shaping galactic structure~\cite{Hu:2000ke, Hui:2016ltb, Mocz:2019pyf}, and DE could be in the form of a quintessential axion~\cite{Kim:2002tq, Ibe:2018ffn, Choi:2021aze}. Efforts towards detecting the signatures of new light particles and their associated fifth forces range from laboratory and space tests~\cite{PhysRevD.50.3614,Hoyle:2004cw,Williams:2004qba,Mota:2006fz,Brax:2007vm,Schlamminger:2007ht,Brax:2011hb,Wagner:2012ui,Burrage:2014oza,Foot:2014osa,TheMADMAXWorkingGroup:2016hpc,Perivolaropoulos:2016ucs,Burrage:2017qrf,Touboul:2017grn,Perivolaropoulos:2019vkb,Blanco:2019hah,Braine:2019fqb,DiLuzio:2020wdo,Bloch:2020uzh,Vagnozzi:2021quy,Tsai:2021lly,Adams:2022pbo,Antypas:2022asj} to cosmological~\cite{Hlozek:2014lca,Baumann:2015rya,DEramo:2018vss,Ade:2018sbj,Poulin:2018dzj,Poulin:2018cxd,Vagnozzi:2019ezj,Vagnozzi:2019kvw,Green:2019glg,EscuderoAbenza:2020cmq,Rogers:2020ltq,Giare:2020vzo,Esteban:2021ozz,DiValentino:2021izs,Vagnozzi:2021gjh,Perivolaropoulos:2021jda} and astrophysical studies~\cite{Jain:2012tn,Arvanitaki:2014wva,Giannotti:2015kwo,Brito:2015oca,Foot:2016wvj,Caputo:2017zqh,Baryakhtar:2017ngi,Stott:2018opm,Roy:2019esk,Davoudiasl:2019nlo,Croon:2020oga,Stott:2020gjj,Desmond:2020nde,Roy:2021uye,Ferlito:2022mok,Chen:2022nbb,Vagnozzi:2022moj,Saha:2022hcd,Tsai:2022jnv}.

In addition, one can also consider modifications to GR  (see e.g.\ \cite{Moffat:2004nw}). In particular, a Yukaka-type potential, referred to in this context as Yukawa gravity~\cite{Borka:2013dba,Zakharov:2016lzv,Dong:2022zvh}, has been adopted in order to parametrize the phenomenological effects of theories with massive gravitons~\cite{deRham:2014zqa}.

The motion of asteroids is continuously and carefully monitored for various reasons that include planetary defense purposes~\cite{2006Sci...314.1276O}, to the extent that dedicated studies have been recently financed both by the National Aeronautics and Space Administration (NASA)~\cite{Gustetic_2018} and the European Space Agency (ESA)~\cite{ESA_Asteroids} asteroid missions. These studies benefit from current and future radar and optical data, including from facilities and missions such as Arecibo (decommissioned), Goldstone, Catalina, the Vera Rubin Observatory (VRO), and Gaia~\cite{2016AJ....152...99N, Djorgovski:2011iy, 2020arXiv200907653V, Carry_2021}. Such studies are not free of challenges, as asteroid trajectories are subject to perturbations that range from gravitational effects from other celestial objects to non-gravitational effects due to the thermal and reflective properties of the asteroid's surface. Recent advances in studying physical parameters and relevant physical processes (including GR parameters, solar quadrupole moment, and Yarkovsky effect) from asteroid data, taking into account these perturbations~\cite{Verma:2017ywb, Greenberg_2020}, inspire us to examine the possibility of probing new physics with asteroid astrometry.

In this paper, we provide a proof-of-principle study using asteroid precessions and astrometric data to probe new ultralight particles. Previously, planets, exoplanets, and Kuiper Belt Objects (KBOs) were used to test GR and/or search for dark sector particles~\cite{Iorio:2005qn, Adler:2008ky,Jordan:2008zi,Iorio:2010rg,Hooper:2011dw,Iorio:2012wv,Overduin:2013soa,2013AstL...39..141P,Iorio:2014yga,Ain:2015mea, Masuda:2016ggi, Blanchet:2019zxv, Bramante:2019fhi,Sun:2019ico,Garani:2019rcb,Scholtz:2019csj, Ruggiero:2020yoq,Chan:2020vsr, Leane:2020wob, Wei:2021xek,Leane:2021tjj}, yet the potential of probing new physics using asteroids remains mostly unharvested. Thanks to advances in radar and optical astrometry, the motion of asteroids, especially those classified as Near-Earth Objects (NEOs), is tracked much more precisely than KBOs and exoplanets. The use of asteroids over planets~\cite{Poddar:2020exe} also carries several advantages, ranging from their sheer number, to their spread in orbital radius allowing to probe a wide range of parameter space. Focusing on light mediators in the mass range $m \simeq (10^{-21}-10^{-15})\,{\rm eV}$, or equivalently in the Compton wavelength range $\lambda \simeq (10^{-3}-10^{3})\,{\rm au}$, we estimate the sensitivity reach of asteroid precessions to the mediator mass and coupling, which we find to be competitive with some of the most stringent equivalence principle tests~\cite{PhysRevD.50.3614, Schlamminger:2007ht, Wagner:2012ui,Berge:2017ovy}, and outline further steps to improve the analysis.
We find that the strongest bounds are realized in the region where $\lambda \sim a$, where $a$ is the asteroid's semi-major axis; this motivates the future use of objects orbiting further away from the Sun to probe lighter mediators efficiently.
While we focus on gauged SM symmetries and baryon-coupled ultralight scalars as concrete examples, our study is broadly applicable to various well-motivated new physics models. 

We expect our study to integrate into ongoing efforts to study gravity in space. We delineate the possibility of conducting similar studies using extended asteroid catalogs, Trans-Neptunian Objects (TNOs), and exoplanets. The growing wealth of available optical and radar data would lead to significant improvements in our results. 

The rest of this paper is then organized as follows. In section~\ref{sec:orbitalprecession} we lay down the theoretical motivations for our study. In section~\ref{sec:methods} we discuss the methods used to assess the observables against asteroid data, with the results presented in section~\ref{sec:results}. Conclusions and future prospects are drawn in section~\ref{sec:conclusions}.

\section{Light particles and orbital precessions}
\label{sec:orbitalprecession}

We consider a celestial body of mass $M_\ast$ orbiting the Sun and subject to an additional fifth-force mediated by a new light particle, with an associated potential $V(r)$ which is modeled as:
\begin{equation}
    V(r) = -\frac{GM_\odot M_\ast}{(1+\alpha_1)r}  \left(1+\alpha_2\,\exp \left ( -\frac{r}{\lambda} \right )\right) \,,
    \label{eq:yukawa}
\end{equation}
where $M_\odot$ is the solar mass and $\lambda$ the Yukawa force range. 
Here, $\alpha_1$ and $\alpha_2$ are Yukawa parameters whose values will be specified later in the text according to the different types of Yukawa models.
The potential in eq.~\eqref{eq:yukawa} leads to deviations from the body's Newtonian orbit, introducing an orbital precession alongside GR effects accounted for by $g_{\mu\nu}$.

We consider planar motion and fix $\theta=\pi/2$. Adopting the inverse radius variable $u \equiv 1/r = u(\varphi)$ and defining $\widetilde{G} \equiv G/(1+\alpha_1)$, we obtain the orbit's fundamental equation (in SI units)~\cite{Poddar:2020exe}:
\begin{equation}
    \frac{{\rm d}^2 u}{{\rm d}\varphi^2} + u - \frac{\widetilde{G}M_\odot}{L^2} \!=\! \frac{3\widetilde{G}M_\odot}{c^2} u^2 \!+\! \alpha_2\frac{\widetilde{G} M_\odot}{L^2}\left(\!1+\frac{1}{\lambda u}\!\right)\!e^{-\frac{1}{\lambda u}}\,,
    \label{eq:full}
\end{equation}
where $L$ is the orbital angular momentum per unit mass. The first term on the right-hand side of eq.~\eqref{eq:full} leads to well-known GR corrections, while the second term leads to fifth force-induced corrections. Solving eq.~\eqref{eq:full} numerically determines the fifth force-induced precession, which, upon comparison to observations, can constrain $\alpha_1$ and $\alpha_2$ given the force range.

For most of the Yukawa potentials generated from new BSM light scalars or vectors, $\alpha_1=0$ and we define $\alpha_2\equiv\widetilde{\alpha}$.
Examples to which our study can be applied include gauged $U(1)_B$~\cite{Carone:1994aa, FileviezPerez:2010gw}, $U(1)_{B-L}$~\cite{Davidson:1978pm,Mohapatra:1980qe,Davidson:1987mh}, $L_{\mu}-L_{e, \tau}$~\cite{Foot:1990mn, He:1991qd,Escudero:2019gzq}, and baryon-coupled scalar~\cite{Blinov:2018vgc,Sibiryakov:2020eir,Izaguirre:2014cza,Pospelov:2017kep} models. In these models: 
\begin{equation}
    \widetilde{\alpha} = \pm \,\frac{g^2}{4\pi G}\frac{Q_\odot\,Q_\ast}{M_\odot M_\ast}\,,
    \label{eq:yukawadarkphoton}
\end{equation}
where + corresponds to attractive and - corresponds to repulsive force under this notation, and $g$ is the coupling strength. For a BSM boson, the mediator mass $m\equiv \hbar c/\lambda$. For illustrative purposes, we shall focus on the mediator being either a gauged $U(1)_B$ dark photon or an ultralight scalar coupled to the baryon number. The coupling is given by $g=g_{\phi, A'}$ for either a scalar ($\phi$) or vector ($A'$) mediator, whose mass is $m = m_{\phi,A'}$. Moreover, $Q_\ast \equiv M_\ast/m_p$ and $Q_\odot \equiv M_\odot/m_p$ are the celestial object and Sun total baryon numbers respectively, with $m_p$ the proton mass. The gauged $U(1)_B$ exhibits a chiral anomaly whose cancellation can be achieved, e.g., by introducing additional appropriately constrained particles~\cite{Dror:2017ehi,Dror:2017nsg} or invoking extra model building~\cite{Green:1984sg,Kaplan:1991dc,Pierce:2018xmy}. In this phenomenological study, we assume no self-interaction for the scalar. We focus on the asteroid phenomenology of these models and we emphasize again that our method is not limited to the $U(1)_B$ dark photon and scalar mediators case studies. For the scalar coupling, we take the Lagrangian to be $\mathcal{L_\phi} \subset (g_{\phi,p}\bar{p}p\:+\:g_{\phi,n} \bar{n}n\:+\:g_{\phi,e} \bar{e}e\:)\phi$, and consider two cases:  
Case (a) would have the same magnitude as the $U(1)_B$ coupling, modulo a different sign (attractive for scalar and repulsive for vectors), so $g_{\phi,p}=g_{\phi,n}$ and $g_{\phi,e}=0$. 

For the case of Yukawa gravity,  $\alpha_1=\alpha_2$ and we define $\alpha_2\equiv \widetilde{\alpha_g}$ to avoid any confusion. We also define the graviton mass $m_g\equiv \hbar c/\lambda$.\\

\section{Methods}
\label{sec:methods}

We specialize to asteroids as the celestial objects of interest. Our goal is to estimate the sensitivity reach for the coupling strength and mediator mass of a Yukawa-type fifth force, using the induced orbital precession. To this end, we focus on nine asteroids with precise radar and optical trajectory determinations, studied in detail in Ref.~\cite{Verma:2017ywb}. These are NEOs with semi-major axes $a \in [0.64,1.08]\,{\rm au}$ and eccentricities $\mathsf{e} \in [0.48 - 0.90]$. We are only interested in the impact of the fifth force field on the induced orbital precession. A fully-fledged analysis entails \textit{a)} computing the fifth force impact on the asteroid trajectory via an appropriate integrator, accounting for perturbations from all nearby objects, and \textit{b)} using raw astrometric measurements of the asteroid's trajectory to constrain the fifth force. As this is the beginning attempt to perform this type of analysis, our aim is simply to estimate the fifth force sensitivity reach, while providing a proof-of-principle for the feasibility of such a study and laying the foundations for future detailed analyses.

Various effects contribute to asteroid orbital precession. Perturbations from planetary motions source the largest contributions to the orbital precession of a body, see e.g.\ Refs.~\cite{2005AmJPh..73..730S, 2012ApJ...757..105K}. Two additional contributors are GR effects and solar oblateness~\cite{Verma:2017ywb}. These two effects contribute to the perihelion precession as measured from a fixed reference direction per orbital period, for an orbit whose inclination angle with respect to the solar equator is $i_{\rm eq}$, as~\cite{Iorio:2008bx}:
\begin{equation}
    \Delta \varphi_0 = \frac{6\pi GM_\odot}{a(1-\mathsf{e}^2)c^2}\left[\frac{2-\beta+2\gamma}{3}\right] + 3\pi R_\odot^2 \frac{2-3\sin^2i_{\rm eq}}{2a^2(1-\mathsf{e}^2)^2}J_2\,,
    \label{eq:precession}
\end{equation}
where $R_\odot$ is the solar radius and $J_2$ the solar quadrupole moment~\cite{Misner:1973prb}. The parameters $\gamma$ and $\beta$ describe the deviations from GR in the Parameterized Post-Newtonian (PPN) approach, with GR being recovered when $\beta = \gamma = 1$~\cite{Will:1972zz, Will:2018bme, Misner:1973prb}, and with deviations from unity being therefrom tightly constrained by Solar System probes~\cite{2001P&SS...49.1445S,Bertotti:2003rm,2012Sci...336..214S}. We verified that the precession cross-contribution from both $J_2$ and $\widetilde{\alpha}$ is sub-dominant. The effects on the apsidal precession coming from the presence of planetary perturbations has been assessed e.g.\ in Ref.~\cite{2018arXiv180207115B}.

To estimate the fifth force sensitivity reach, we impose that the new physics contribution to the orbital precession in eq.~\eqref{eq:deltaphi} does not exceed the uncertainty budget associated to the two major precession contributors $\beta$ and $J_2$ (as $\vert 1-\gamma \vert$ is more tightly constrained than $\vert 1-\beta \vert$). We lean upon the results of Ref.~\cite{Verma:2017ywb}, who estimated the sensitivity to $\beta$ and $J_2$ obtainable from a fully-fledged analysis of the 9 asteroids: the analysis was based on the Mission Operations and Navigation Toolkit Environment (MONTE) software~\cite{2018CEAS...10...79E}, which numerically integrates the orbit equations of motion, using a dynamical model including gravitational perturbations from nearby celestial objects and accounting for Yarkovsky drift.\footnote{Newton's constant $G$ is measured independently in cold-atom experiments and other techniques to a relative uncertainty of $\sim 2\times 10^{-5}$~\cite{Rosi:2014kva, Hamilton:2015zga, Xue:2020spa}.}

We obtain the precession $\Delta \varphi$ by numerically solving eq.~\eqref{eq:full} with initial conditions $u(0) = [a(1-\mathsf{e})]^{-1}$\ and $u'(0) = 0$, corresponding to an elliptic orbit with eccentricity $\mathsf{e}$ and at its perihelion for $\varphi = 0$. The induced precession is estimated by expressing $u=[a(1-\mathsf{e}^2)]^{-1} [1+\mathsf{e}\cos\varphi (1-\delta)]$, solving for $\delta$, and deriving the precession as $\Delta\varphi = 2\pi\delta/(1-\delta) $. The new physics contribution is then 
$\Delta \varphi_{\rm Yukawa}(\alpha_1, \alpha_2, \lambda)=\Delta \varphi-\Delta \varphi_{\rm GR}$.

The very light mediator limit $m_\phi \ll \hbar/ac$ (or, equivalently, $\lambda \gg a$) admits an analytical expression for the fifth force-induced precession, obtained by expanding around the exponential term:
\begin{align}
    \vert \Delta \varphi_{\rm Yukawa} \vert \simeq  \frac{2\pi\widetilde{\alpha_2}}{1+\widetilde{\alpha_2}} \left ( \frac{am_\phi c}{\hbar} \right ) ^2 \left ( 1-\mathsf{e} \right ) = \frac{2\pi\widetilde{\alpha_2}}{1+\widetilde{\alpha_2}} \left ( \frac{a}{\lambda} \right ) ^2 \left ( 1-\mathsf{e} \right ).
    \label{eq:deltaphi}
\end{align}
We stress that we \textit{do not} use this approximation to estimate our sensitivity reach, but numerically solve eq.~\eqref{eq:full}, later verifying that eq.~\eqref{eq:deltaphi} holds when $\lambda \gg a$. Note that the precession goes to zero in the limit $\lambda \to \infty$ where the Newtonian $1/r$ potential is recovered. In this limit $|\Delta \varphi_{\rm Yukawa}|\propto a^2$, which carries a different functional dependence on $a$ compared to the GR and $J_2$ contributions, so studying objects within a wide range of $a$ and $\mathsf{e}$ can help differentiate the contributions from these terms.

Note that $ \! \left|\frac{\partial \Delta \varphi_0}{\partial \beta}\right|\sigma_\beta \!\sim\! \left|\frac{\partial \Delta \varphi_0}{\partial J_2}\right|\sigma_{J_2}$, meaning that both parameters are determined to comparable levels as far as precessions are concerned, and we want to estimate the range of uncertainty for the Yukawa parameters at a given $\lambda$. We therefore require that the new physics precession contribution does not exceed the uncertainty budget associated to $\beta$ and $J_2$:
\begin{equation}
    \Delta \varphi_{\rm Yukawa}^2 \!<\! \left|\frac{\partial \Delta \varphi_0}{\partial \beta}\right|^2\!\!\sigma_\beta^2 \!+\! \left|\frac{\partial \Delta \varphi_0}{\partial J_2}\right|^2\!\!\sigma_{J_2}^2 \!+\! 2\rho\!\left|\frac{\partial \Delta \varphi_0}{\partial \beta}\,\frac{\partial \Delta \varphi_0}{\partial J_2}\right|\!\sigma_{J_2}\sigma_\beta\,,
    \label{eq:demand}
\end{equation}
where $\rho$ is the correlation coefficient. The above inequality is a function of the fifth force parameters, and values thereof which saturate the inequality give our estimated sensitivity reach. We repeat these steps for each of the 9 asteroids, obtaining 9 separate (but comparable) limits  on the $\alpha$'s for different values of $\lambda$. We implicitly assume that the central values of the measured orbital precessions are consistent with the expectations given GR and all nearby perturbers modelled in Ref.~\cite{Verma:2017ywb}, and therefore that there is no detection of fifth force, whose contribution accordingly must not exceed the precession uncertainty budget. In other words, our analysis is akin to a forecast around a fiducial model with no fifth force. For this reason, we also do not account for the perturbations due to the planetary motion in the Solar System, since the presence of a fifth force could already be altering the results from the tinier contributions from GR effects and solar oblateness.

A covariance analysis of the 9 asteroids based on a 2022 sensitivity projection infers $\sigma_\beta = 5.6 \times 10^{-4}$ and $\sigma_{J_2} = 2.7 \times 10^{-8}$, and a correlation coefficient $\rho = -0.72$, with a Monte Carlo forecast recovering similar figures~\cite{Verma:2017ywb}. We base our sensitivity reach estimate on the 2022 values to reflect the current sensitivity. We also present an estimate based on the ``optimistic'' 2022 values $\sigma_\beta=2\times 10^{-4}$ and $\sigma_{J_2}= 10^{-8}$ given in Ref.~\cite{Verma:2017ywb}.

\section{Results and Discussion}
\label{sec:results}

\begin{figure*}
    \centering
    \includegraphics[width=0.49
\textwidth]{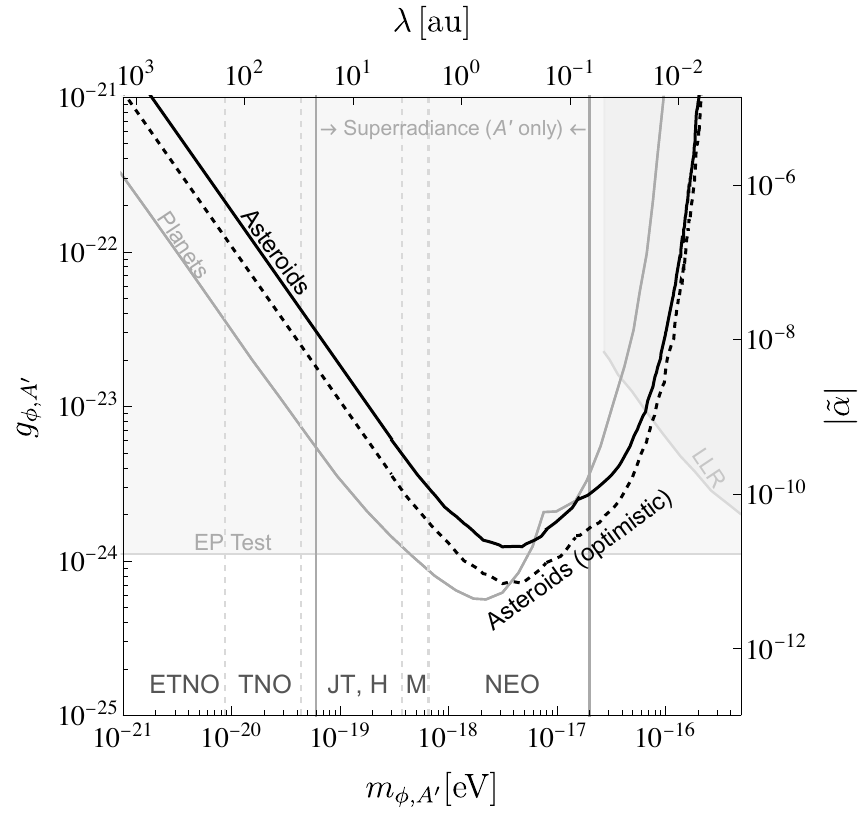} \qquad
\includegraphics[width=0.45\textwidth]{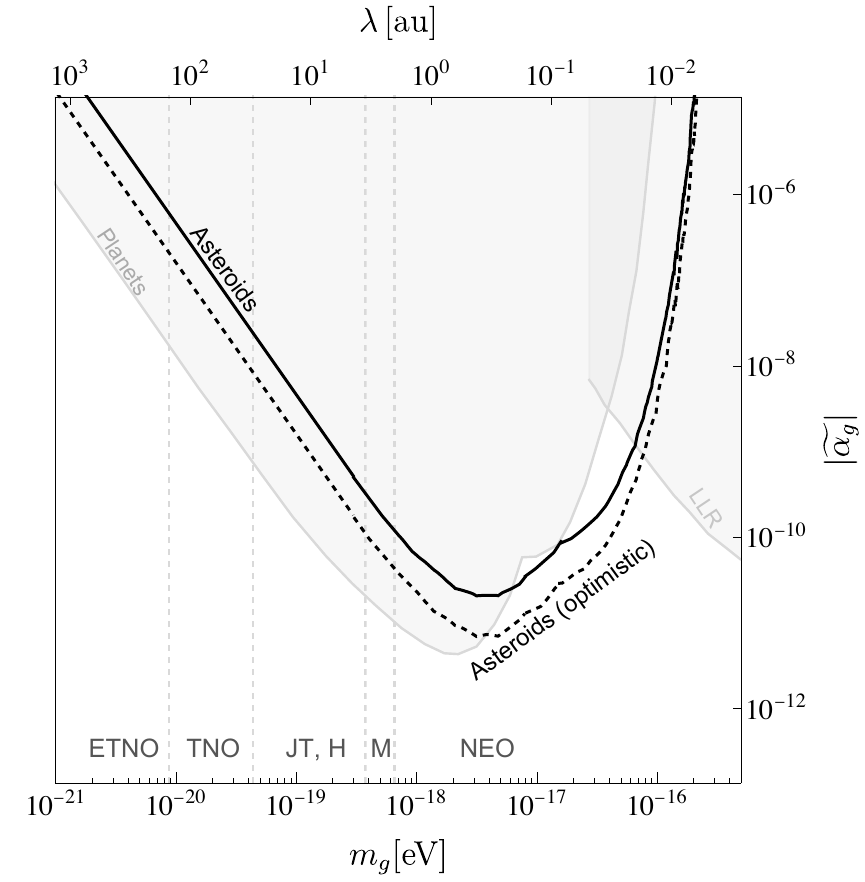}
\caption{Estimated sensitivity reach for the mass and coupling strength of {\bf (a)} $U(1)_B$ dark photons and baryon-coupled scalars with $g_{\phi,p}=g_{\phi,n}$ and $g_{\phi,e}=0$  {\bf(left panel)}; {\bf(b)} Yukawa gravity with modification parameter $\widetilde{\alpha_g}$ as a function of the graviton mass $m_g$ {\bf(right panel)}. The asteroid curves are obtained by studying the precessions of nine NEO asteroids with $a \in [0.64,1.08]\,$au and $\mathsf{e} \in [0.48 - 0.90]$. The solid black curve shows the tightest $1\sigma$ sensitivity reaches from asteroids TU3, MN, and BD19, while the dashed black curve is the $1\sigma$ sensitivity reach based on the optimistic 2022 projection of Ref.~\cite{Verma:2017ywb}. Existing constraints include those from planets~\cite{Poddar:2020exe}, EP tests~\cite{Berge:2017ovy}, and vector superradiance~\cite{Baryakhtar:2017ngi}, the latter is only applicable to the dark photon $A'$. LLR~\cite{Williams:2004qba} provides the leading bound for masses $\gtrsim 10^{-16}\,$eV and is included for completeness. As a note of caution, while laboratory bounds are shown along with the results from our analysis, the length scales probed by the two different techniques differ by orders of magnitude. They provide complementary probes of the fifth forces with different screening effects.}
    \label{fig:constraints2}
\end{figure*}

In figure~\ref{fig:constraints2} we show the estimated sensitivity to the $U(1)_B$ dark photon and baryon-coupled ultralight scalar couplings, as a function of their masses. In these examples, all the baryons in the Sun and the asteroids are charged and the specific compositions thereof do not affect our results. Three specific asteroids, i.e.\ TU3, MN, and BD19, deliver the strongest limits (see figure~\ref{fig:constraints} in the Appendix for the sensitivity reaches for each of the asteroids), given by the solid curve in the figure (whereas the dashed curve shows a stronger sensitivity reach based on the optimistic projection described earlier). We have chosen to report the \textit{tightest} sensitivity reach since all 9 curves are comparable.
Unsurprisingly, the peak sensitivity is achieved for mediator masses approximately corresponding to the (inverse) orbital radius of each asteroid. On the same figure, we also mark the regions corresponding to typical orbital radii of other (non-NEO) asteroids and TNOs. Finally, we note that the inferred sensitivity to the fifth force coupling strength and mediator mass within the $U(1)_B$ model can easily be converted to other long-range forces, including those associated to gauged $U(1)_{B-L}$ and $L_{e}-L_{\mu, \tau}$ symmetries, following Refs.~\cite{Adelberger:2006dh,Fayet:2018cjy}.

For a fixed coupling strength and mediator mass, the repulsive (attractive) force from a vector dark photon (scalar) mediator would yield a precession of essentially equal magnitude, as we have checked numerically, resulting in eq.~\eqref{eq:demand} delivering identical sensitivity reaches. A fully-fledged analysis of the raw asteroid astrometric data should account for the different sign in the precession contribution, and might therefore return different constraints for the two cases.

Also shown in figure~\ref{fig:constraints2} are existing leading bounds from equivalence principle tests~\cite{PhysRevD.50.3614,Schlamminger:2007ht,Wagner:2012ui,Berge:2017ovy}, black hole (BH) superradiance~\cite{Baryakhtar:2020gao}, and planetary precession~\cite{Poddar:2020exe}. Lunar Laser Ranging (LLR)~\cite{Williams:2004qba,Murphy:2013qya} provides the leading bound for masses $\gtrsim 10^{-16}\,$eV and is included for completeness. A concept for a hypothetical space mission similar to LLR exploiting the Martian moon Phobos is also under development, and is referred to as Phobos Laser Ranging (PLR)~\cite{Turyshev:2010gk}. Such a test would be extremely sensitive to the parameter $\widetilde\alpha$ and could potentially improve bounds by two orders of magnitude on scales of an astronomical unit. Only the vector superradiance bound is present since the scalar superradiance one requires further studies on supermassive BHs, owing to large uncertainties concerning their environments. Asteroids offer an actual probe of fifth forces with range beyond the au scale that is complementary to the scales probed within laboratories; more complex long-range force models may be invoked to bypass torsion balance constraints~\cite{Khoury:2003rn,Burrage:2016bwy}. We have also set a new limit on Yukawa gravity.

Furthermore, we show in figure~\ref{fig:constraints2} that precession tests, including asteroid ones, are potentially competitive with EP laboratory tests. This demonstrates that precession tests from asteroids and other planetary objects are especially suitable in probing EP-conserving (or approximately conserving) long-range forces.\footnote{Satellite-based EP tests have recently reported no evidence for EP violation~\cite{MICROSCOPE:2022doy}, or are planned for deployment such as the Satellite Test of the Equivalence Principle (STEP)~\cite{Overduin:2012uk}.} 
Also note that, while laboratory bounds are shown along with the results from our analysis, the length scales probed by the two different techniques differ by orders of magnitude. They provide complementary probes of the fifth forces with different screening effects.
We have checked that the results obtained are consistent with the requirement that the precession contribution from the fifth-force is below the limits from GR. This can be seen with an explicit computation at small scales $\lambda \lesssim a$ by comparing the results in Eq.~\eqref{eq:deltaphi} with the precession expected in GR, leading to:
\begin{equation}
    g_{\phi,A'} \lesssim 6\times 10^{-23}\,\frac{1}{(1-\mathsf{e})(1+\mathsf{e})^{1/2}}\,\left(\frac{a}{\rm au}\right)^{-3/2}\,\left(\frac{m_{\phi,A'}}{10^{-18}{\rm\,eV}}\right)^{-1}\,,
\end{equation}
a much looser requirement than the bounds derived in Fig.~\ref{fig:constraints2} for all asteroids considered.

Summing up, we note that prospects for advancing our understanding of these nine asteroids, including their hazardous or complex nature, are bright. For example, the binary-asteroid system (66391) 1999 KW4/Moshup is a potential threat to Earth due to its orbital trajectory, and is the subject of intense studies~\cite{1999KW4}.

\section{Conclusions and Future Prospects}
\label{sec:conclusions}

Our work attempts to connect fundamental new physics and astrometry data for planetary objects. Focusing on nine near-Earth asteroids, our analysis provides a general recipe and sensitivity reach estimate for long-range fifth forces induced by ultralight mediators. Follow-up opportunities are detailed below.

{\it New target objects ---} There are opportunities to extend our study to ${\cal O}(10^6)$ minor planets, classified in table~\ref{tab:targets}. Including asteroids and comets, there are $\sim 25000$ NEOs (comets are less ideal for our study since they are subject to strong non-gravitational perturbations), a significant number of which have orbits that can be tracked to a similar level of precision as the nine asteroids considered. Among these nine asteroids, the one whose trajectory is determined to lowest accuracy is 2004 KH17, whose semi-major axis is nonetheless measured to $\simeq 1\,{\rm km}$ precision. Currently $\sim 1800$ NEOs have orbits known to comparable or higher accuracy: 247 of these has been analyzed to study Yarkovsky drift~\cite{Greenberg_2020}. Neglecting systematics, $\sim 1800$ NEOs can potentially improve our sensitivity reach by more than 1 order of magnitude.

Beyond NEOs, other asteroids including main-belt asteroids (M), Hildas (H), and Jupiter Trojans (JT) can serve similar purposes. Their larger semi-major axes imply that their sensitivity reaches would peak at lower mediator masses, allowing us to probe lighter dark sector particles. Moreover, as the Yarkovsky drift weakens with increasing distance from the Sun as $a^{-1/2}$~\cite{2017AJ....153..108G}, the effects over Hildas and Trojans would be negligible compared to NEOs for kilometer-sized bodies and for the same time of the observations. Achieving precision comparable to NEOs might be challenging, but spacecraft ranging may provide data with precision rivalling/surpassing radar observations: for example, the LUCY space mission~\cite{2021PSJ.....2..172O} will provide precision data for Trojans.

TNOs and ETNOs, residing in the outer Solar System, are of extreme interest owing to their trajectories being subject to significantly less gravitational perturbations and solar thermal effects. Their large semi-major axes mean they can be used to probe ultra-light mediators at even lower masses. All these objects are labelled in figure~\ref{fig:constraints2} according to their typical semi-major axes.
\begin{table}[t!]
    \def\arraystretch{1.5}
    \centering
    \begin{tabular}{lll}
    \hline\hline
    Minor Planets \hspace{0.5cm} & $a\,$[au] \hspace{0.3cm} & $\sim$ Numbers \hspace{0.8cm} \\
    \hline
    Near-Earth Objects (NEOs) &  $< 1.3^*$ & $>25000$ \\
    Main-Belt Asteroids (M) & $\sim 2-3$ & $\sim$ 1 million \\
    Hildas (H) & 3.7 - 4.2 & $>4000$ \\
    Jupiter Trojans (JT) & 5.2  & $>9800$ \\
    Trans-Neptunian Objects (TNOs) & $>30$ & 2700 \\
    Extreme TNOs (ETNOs) & $>150$ & 12 \\
    \end{tabular}
    \caption{Targets for our future studies, for which opportunities are provided by sheer numbers and observational programs, classified roughly based on their typical semi-major axes.\\ $^*$NEOs are defined as having perihelia $a(1-\mathsf{e})<1.3\,{\rm au}$.}
    \label{tab:targets}
\end{table}

{\it New observations ---} Radar studies including Goldstone and the recently decommissioned Arecibo~\cite{2016AJ....152...99N} have been collecting high-precision NEOs astrometrical data. VRO will discover a factor of 5 more Solar System minor objects (see table~1 of Ref.~\cite{2020arXiv200907653V}), while other optical sky surveys such as Catalina~\cite{Djorgovski:2011iy}, Pan-STARRS~\cite{2013PASP..125..357D}, ATLAS~\cite{2018PASP..130f4505T}, DECam~\cite{2021arXiv210403411G}, and ZTF~\cite{Graham_2019} will also be of great use to such studies.  High precision astrometry is also achievable with space-based telescopes such as Hubble~\cite{2020arXiv201015425P}, James Webb~\cite{Rivkin_2016}, Euclid~\cite{Carry_2018}, and Roman~\cite{2019arXiv190205569A}. LUCY~\cite{2021PSJ.....2..172O} will visit Trojans and the JANUS spacecraft will investigate two binary asteroids~\cite{2021arXiv210302435L, Seligman_2021}, providing valuable information to extend our study. New astrometrical techniques such as occultation can substantially improve orbital trajectory determinations~\cite{2010arXiv1001.2010W}. Asteroids also affect gravitational wave detections through gravity gradient noise~\cite{Fedderke:2020yfy,Fedderke:2021kuy}.

{\it Data storage and dedicated software development ---} Our work motivates the study and inclusion of precession measurements for objects stored in the JPL small objects~\cite{nasa} and Minor Planet Center~\cite{iau} databases. On the analysis side, a fully-fledged study entails re-analyzing the (raw) astrometric asteroid trajectory data. Dedicated computing platforms such as MONTE~\cite{2018CEAS...10...79E}, self-consistently modelling all relevant physical effects, can be used to this end after appropriate modification to include the fifth force effect. We expect this to be an important task for future studies~\cite{precite}.

{\it Theory ---} Our study can be viewed as an investigation of a specific example of deviations from GR and/or the SM. Of course, the method can be extended to test other theories of gravity (e.g.~\cite{Bergmann:1968ve,1983ApJ...270..365M,Sotiriou:2008rp,Clifton:2011jh,Nojiri:2017ncd}), or other types of dark sector models~\cite{Kaplan:2009de,Farzan:2012hh,Cyr-Racine:2012tfp,Petraki:2014uza,Randall:2014lxa, Foot:2014uba,Tsai:2020vpi, Knapen:2017xzo, Blinov:2018vgc}, by computing their effects on the dynamics of celestial objects~\cite{precite}. One can also consider asteroid tracking arrays (ATAs), analogously to pulsar timing arrays, to study gravitational waves and other aspects of fundamental physics.

{\it Final outlook ---} We expect to broaden up attempts at probing fundamental physics from astrometric data for minor planets in the inner and outer Solar System. More generally, alongside seminal works~\cite{Iorio:2005qn,Adler:2008ky,Jordan:2008zi,Iorio:2010rg,Hooper:2011dw,Iorio:2012wv,Overduin:2013soa,Iorio:2014yga,Ain:2015mea,Masuda:2016ggi,Blanchet:2019zxv,Bramante:2019fhi,Sun:2019ico,Garani:2019rcb,Scholtz:2019csj,Ruggiero:2020yoq,Chan:2020vsr,Leane:2020wob,Wei:2021xek,Leane:2021tjj}, we have only just begun exploring the full potential of establishing connections between microscopic new physics and macroscopic planetary observations, from near (NEOs) to far (exoplanets) celestial objects.

\section*{Acknowledgments}
We thank Alex Drlica-Wagner, Davide Farnocchia, Adam Greenberg, Nick Gnedin, Marco Micheli, Olivier Minazzoli, Matthew Payne, Darryl Seligman, Leo Stein, and Quanzhi Ye for useful discussions regarding asteroid studies, and astronomy and planetary observations in general. We also thank Masha Baryakhtar, Nikita Blinov, Cedric Delaunay, Robert Lasenby, Tanmay Kumar Poddar, Tracy Slatyer, Yotam Soreq, Liantao Wang, Yue Zhang, and Yue Zhao for discussions regarding ultralight dark sector studies.  We are grateful to Yuval Grossman, Marco Micheli, and Darryl Seligman for their invaluable comments on our draft. This document was prepared by Y-D.T.\ using the resources of the Fermi National Accelerator Laboratory (Fermilab), a U.S.\ Department of Energy, Office of Science, HEP User Facility. Fermilab is managed by the Fermi Research Alliance, LLC (FRA), acting under Contract No.~DE-AC02-07CH11359. Y-D.T.\ also thanks KICP, University of Chicago, for hospitality. Y.W.\ acknowledges support from an MICDE Catalyst grant at University of Michigan, DoE grant DE- SC007859, and the LCTP at the University of Michigan.S.V. was partially supported by the Isaac Newton Trust and the Kavli Foundation through a Newton-Kavli Fellowship, and by a grant from the Foundation Blanceflor Boncompagni Ludovisi, n\'{e}e Bildt.

\section{Appendix}
\label{sec:appendix}

\begin{figure}[ht]
    \centering
    \includegraphics[width=0.7\linewidth]{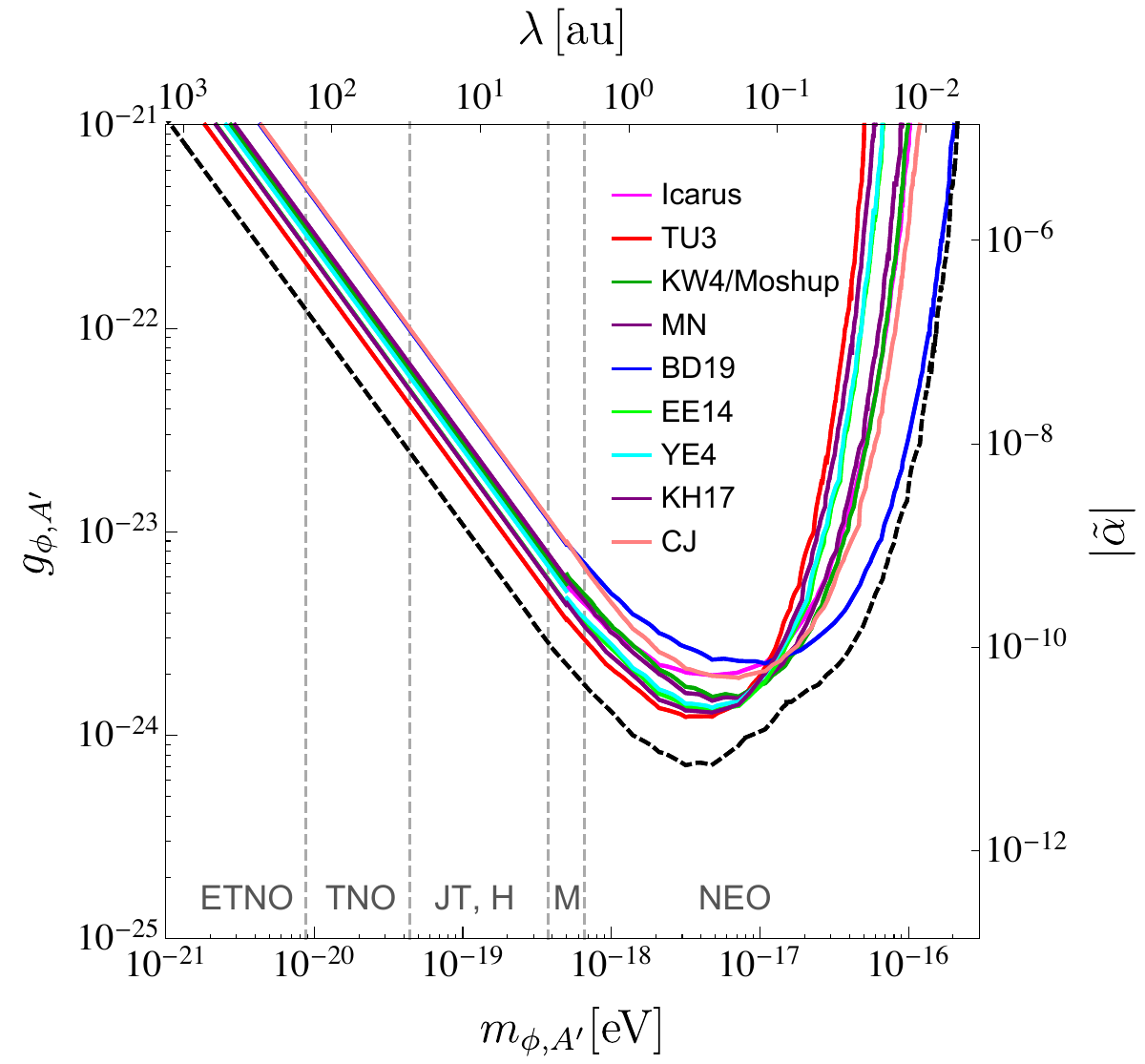}
    \caption{Estimated sensitivity reach from each of the 9 asteroids (solid colored curves), and leading sensitivity reach based on the optimistic projection in Ref.~\cite{Verma:2017ywb} (dashed black curve). Asteroids are new probes of long-range forces in the $\sim\,{\rm au}$ range. The sensitivity can be improved by investigating an additional $\sim 25000$ NEOs. Long-range forces at larger distances can be studied using main-belt asteroids (M), Jupiter Trojans (JT), Hildas (H), TNOs, and ETNOs, as discussed in the main text.}
    \label{fig:constraints}
\end{figure}

The method we proposed leads to an estimate for the sensitivity reach for each of the nine individual asteroids. This is shown in more detail in figure~\ref{fig:constraints} in the mass-coupling plane, together with the leading sensitivity reach based on the optimistic projection in Ref.~\cite{Verma:2017ywb} (dashed black curve). Results are comparable across all asteroids.

\bibliographystyle{JHEP}
\bibliography{asteroids.bib}

\end{document}